# Current-Driven Domain Wall Motion: Velocity, Current and Phase Transition

Hao Yu*

Department of Mathematical Sciences, Xi'an-Jiaotong Liverpool University, Suzhou, Jiangsu, 215123, China

[Abstract] The relation between domain wall motion and intensity of driven current is examined in a phenomenological theory where the kinetic energy is expanded as a series of polynomial function of current density just as the Landau phase transition theory. The dependency of velocity on current density is root square which degenerates into linear if the current is much higher than the critical value. The theory result is consistent with several previous experiments and also can explain the change of critical current in the presence of temperature. The role of temperature plays in the dynamics of domain wall motion is also discussed. The phase transition theory in terms of current density is employed to explain the critical behavior of domain wall motion.

The dynamics of domain wall (DW) motion in ferromagnetic materials and devices is essential and has been extensively studied since magnetic data storage technology emerged [1]. The manipulation of magnetic domain or spin by injection of electric current instead of magnetic field is an alternative approach to read-write head in magnetic storage. The relation between velocity of DW and density of current injected is crucial to DW motion dynamics in a so-called spin transfer torque effect [2, 3]. DW can be shifted by polarized electric current due to spin transfer torque [4]. Efforts have been made in both theoretical and experimental research to explore the mechanism of domain wall motion induced by current[5-8]. To reduce the high density of critical current (normally $>10^6$ A/cm$^2$), the threshold to de-pinning domain wall, is a challenge to industry application. Moreover, the velocity of DW is significant to read-write speed of storage devices and the velocity function with respect to current density is also an interesting theme in spintronics.

A direct observation of current-induced DW displacement was reported [5] in which the velocity of DW as a function of current density and duration was quantitatively discussed. DW speed increases with the current density, consistent with the spin-transfer torque mechanism. In consequent theoretical work [6] momentum transfer and spin transfer are compared, and the latter contributes to the DW dynamics in magnetic metal wire, where there is a threshold intensity of current ($j_s^{cr}$) determined by the anisotropy. Thus the wall velocity can be expressed as the function of spin current ($j_s$), $<v> = \sqrt{j_s^2 - j_s^{cr2}}$. Another theory [7] considering current-spin-coupling Hamiltonian and Landau-Lifshitz-Gilbert (LLG) equation states that the average velocity of DW is linearly proportional to current. Further experiment in ferromagnetic semiconductor (GaMn)As [8] show linear or square root dependencies of velocity and current as well as a scaling law of DW creep at various temperatures.

In this article, a phenomenological theory is applied to explain several experimental results [5,8] of current-driven DW displacement to show the functional relationship between velocity and current density.

A moving DW has kinetic energy, and it is proportional to the square of velocity *v* of DW if we define an effective mass $M_w$ of DW, where $M_w$ is not the real mass of the DW region of a magnetic materials but to stand for the inertia of DW, i.e. how difficult to make it move. Therefore the kinetic energy, or the free energy of a DW when it is driven to move is $K = \frac{1}{2}M_w v^2$. The free energy is also the function of current density and temperature, as expected by the theory of spin angular momentum transfer. If temperature is constant, current density *j* is only variable. Both experiments and theories show that there is a threshold or critical current $j_c$. The free energy can be expanded in Taylor series as the function of $j$ around $j_c$, as what Landau did,

$$K = K_0 + A(j-j_c) + B(j-j_c)^2 + \ldots, \qquad (1)$$

where it is assumed that the higher order (more than 2) terms are small and can be omitted. $A$ and $B$ are positive constants related to temperature. So $v$ is the function of $j$ as

$$v = \sqrt{\frac{2}{M_w}[K_0 + A(j-j_c) + B(j-j_c)^2]}. \qquad (2)$$

When $j$ is close to $j_c$, i.e. $j - j_c \gg (j-j_c)^2$, we have $v \propto \sqrt{2A(j-j_c)/M_w}$, and there is a square root dependency of $v$ on $j$. However if the current density is much more than the critical one, i.e. $j - j_c \ll (j-j_c)^2$, then $v \propto \sqrt{2B/M_w}(j-j_c)$ and $v$ is linearly proportional to $j$. In other words, if the current density is much bigger, $v$ and $j$ is nearly linear. These two cases can explain the different results in two theoretical works [6] and [7].

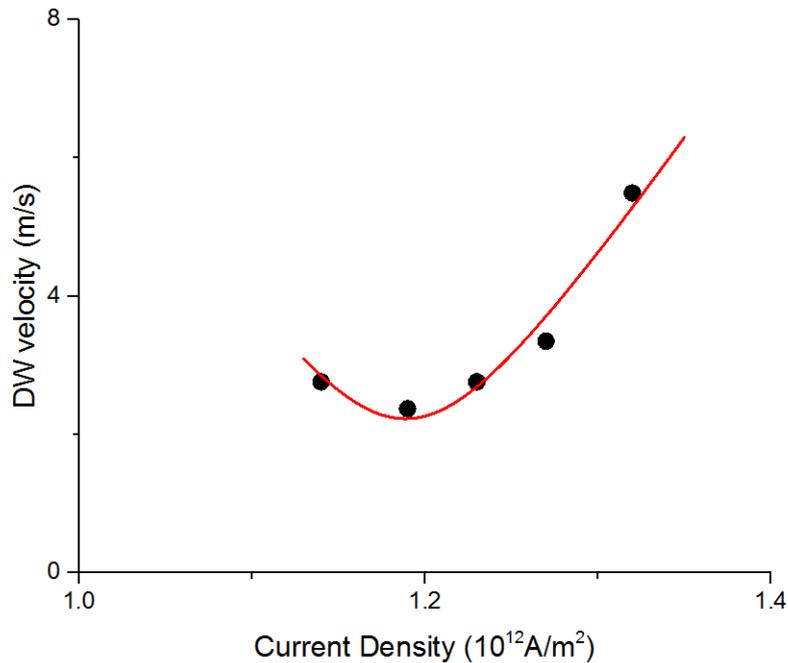

FIG. 1 DW velocity as a function of current density, red curve from our theory prediction well fits the black dots are experimental data taken from[5].

The function is applied to fit the experimental data from Yamaguchi, et al [5]. They show that the average speed of DW increases with the current but the relationship is not a simple linear one. It is interesting to see that when $j>j_c$, there is a minimum of *v* in experiment, which can be demonstrated in our theory. If there were more experimental data, we may calculate out the exact value of the effective mass of domain wall $M_w$, and the coefficient *A* and *B* related to temperature.

The role of temperature can be evaluated by adding thermal terms. The coefficient A and B could be A(*T*) and B (*T*) when temperature *T* varies, which can be expanded in series and only the first order term is taken into account to simplify the model, and consequently A(*T*) and B (*T*) are linearly proportional to *T*: A(*T*)=*aT* and B(*T*)=*bT*. So we have

$$v = \sqrt{\frac{2}{m^*}[K_0 + aT(j-j_c) + bT(j-j_c)^2]}. \qquad (3)$$

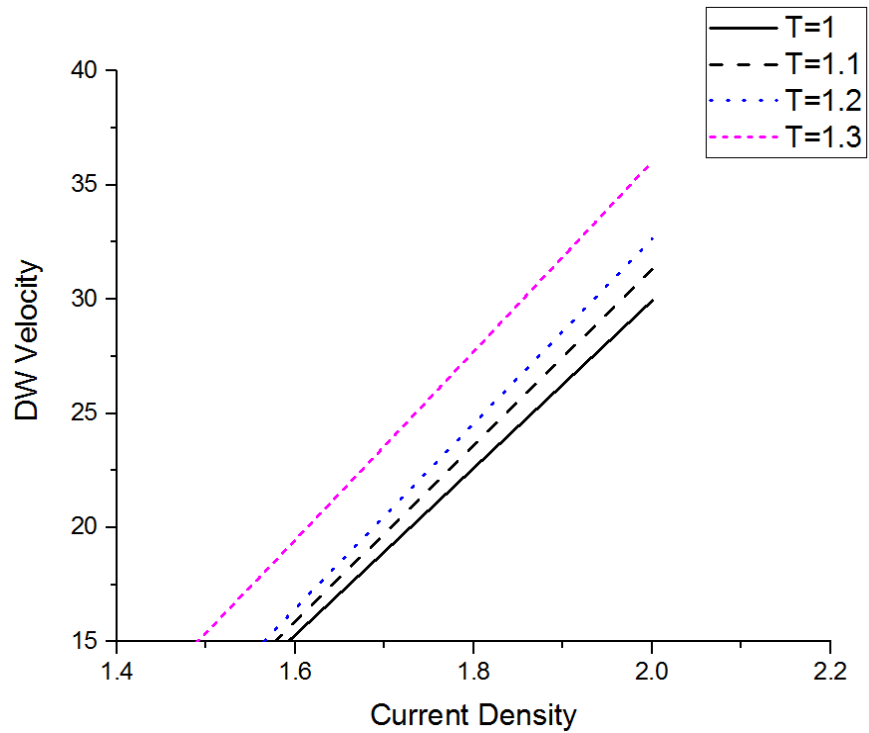

FIG 2. DW velocity as function of current density (higher than critical current) at various temperatures.

The DW velocity as function of current density at various temperatures can be calculated through Eq. (3), shown in Fig. 2. If the temperature T increases, the velocity curve shifts up. The critical intensity of current decreases with increasing temperature may be due to thermal activation. It can be concluded that high temperature helps to effectively reduce the threshold current density. The critical current density as a function of temperature can be obtained in terms of fitting the experimental curve. A more detailed analysis according to the impact of temperature on DW motion needs the function to be expanded to higher order.

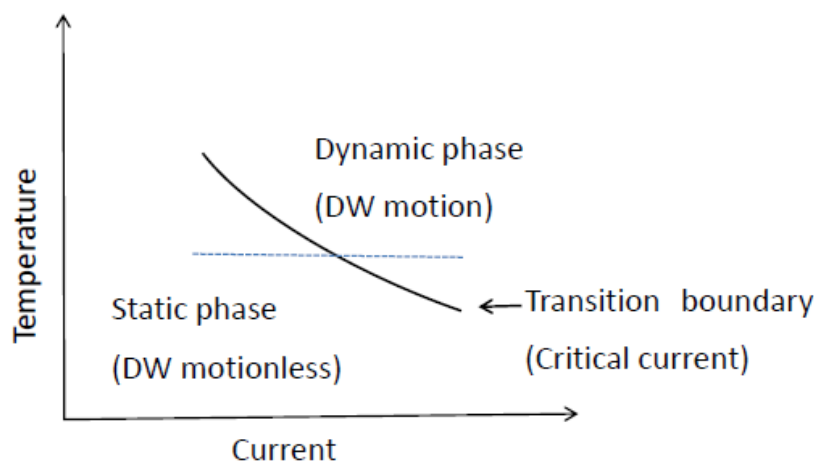

FIG. 3 The phase diagram of DW motion. There are two phases separated by a transition boundary. The dashed line indicates a route in which increases current but fix temperature.

Since the above analysis is based on a calculation like free energy expansion of Landau theory, the existence of critical intensity of current indicates that the DW motion induced by electrical current can be employed a phase transition theory. FIG. 3 represents the phase diagram of a DW motion. There is a boundary between dynamic phase and static phase. Moves right along the dashed line, namely keep the temperature constant, the DW will displaces only when cross the boundary into the dynamic phase. Below the critical boundary, DW remains stationary. The critical point in this 'phase transition' is helpful for us to find a potential approach to reduce the critical intensity of current which is the main challenge to the application of current-driven spin transfer devices.

In conclusion, a phenomenological theory is demonstrated to investigate the relation between DW motion and current density in the absence or presence of temperature. The dependency of velocity of DW on current is square root and degenerates to linear in the case of high density of current. The theoretical calculation is consistent with experimental measurement on DW motion in fine metal or semiconductor wires. Higher temperature is essential to decrease the threshold current intensity above which DW is activated to move. Finally the phase diagram of DW motion is sketched for understanding the dynamics of DW and it may be useful for finding a route to effectively reduce the critical density of electrical current and power dissipation in spin transfer devices.

This work was supported by National Natural Science Foundation of China (No. 11204245).

____________________________________

*Email: hao.yu@xjtlu.edu.cn

References

[1] L. Berger, J. Appl. Phys. 55, 1954 (1984).

[2] L. Berger, J. Appl. Phys. **71**, 2721 (1992).

[3] J. C. Slonczewski, J. Magn. Magn. Mater. **159**, L1 (1996).

[4] L. Berger, Phys. Rev. B **54**, 9353 (1996).

[5] M. Yamaguchi, T. Ono, S. Nasu, K. Miyake, K. Mibu, and T. Shinjo, Phys. Rev. Lett. **92**, 077205 (2004).

[6] G. Tatara and H. Kohno, Phys. Rev. Lett. **92**, 086601 (2004).

[7] S. E. Barnes and S. Maekawa, Phys. Rev. Lett. **95**, 107204 (2005).

[8] M. Yamanouchi, D. Chiba, F. Matsukura, T. Dietl and H. Ohno, Phys. Rev. Lett. **96**, 096601 (2006).